# Data Sustainability and Reuse Pathways of Natural Resources and Environmental Scientists

Yi Shen, Ph.D.
Associate Professor and Research Environments Librarian
Virginia Polytechnic Institute and State University
Blacksburg, Virginia 24061, USA.
https://johnshopkins.academia.edu/YiShen

# Abstract

This paper presents a multifarious examination of natural resources and environmental scientists' adventures navigating the policy change towards open access and cultural shift in data management, sharing, and reuse. Situated in the institutional context of Virginia Tech, a focus group and multiple individual interviews were conducted exploring the domain scientists' all-around experiences, performances, and perspectives on their collection, adoption, integration, preservation, and management of data. The results reveal the scientists' struggles, concerns, and barriers encountered, as well as their shared values, beliefs, passions, and aspirations when working with data. Based on these findings, this study provides suggestions on data modeling and knowledge representation strategies to support the long-term viability, stewardship, accessibility, and sustainability of scientific data. It also discusses the art of curation as creative scholarship and new opportunities for data librarians and information professionals to mobilize the data revolution.

# 1. Introduction

With the enhancement of sensor technology and proliferation of unmanned aircraft applications, the amount of natural resources and environmental science data is growing exponentially. The sheer amount of data, its heterogeneity and complexity, necessitates the use of state-of-the-art data handling techniques along with heterogeneous data interlinking, combination, and integration. Beyond large computing power and storage capacity, data quality, availability, and usability are at the core of scientific research aiming to discover patterns in large amounts of data and to convert these patterns into usable information and exchangeable knowledge. In this regard, provenance of content, transparency of analytics, and quality of information together define and ensure the trustworthiness and usefulness of Big Data. On top of these factors, data ontology and information structuring further enable knowledge sharing by both building a machine-readable set of standard definitions for a scientific domain, and then creating a taxonomy of classes, subclasses, and relationships between them. To harness the immense power of data in the fast-developing field of natural resources and environmental science, rigorous management and robust representation is vastly needed.

New environments require new skills. In tandem with their research, scientists are expected to contribute to the management and organization of data from their scientific work. With these new demanding tasks, it is important to study how research practitioners understand the changing nature of data work and deal with the shifting culture of data practices. In particular, this study investigated the perspectives of scientists whose research centers on forest resources and environmental conservation. Based at a nationally top-ranked natural resources program at Virginia Tech, these scientists approach critical resources and sustainability challenges from multi-disciplinary angles and with diverse data sources, types, and forms. By examining the emerging dynamics and practical challenges of scientists' data work, this study contributes to user experience design and scientific data curation for long-term reuse.

Human-centered data practice studies are not only useful for developing reuse pathways for researchers, but also helpful in defining professional roadmaps for data librarians and information specialists. With the many layered abstractions and materializations of data, information, and knowledge, the fields of data science and information studies are grappling with socio-technical reconfigurations of a widening scope and scale. "To address the complex, mediating requirements of making data work" (International Data Week, 2016), data librarians and information agents not only need the skills to support the technical requirements and production needs for metadata handling, but also need sound skills in data analysis, remediation, and normalization. Moreover, they need the ability to develop and/or customize applications and tools for automation and innovation.

To substantiate and contextualize scientists' data work experiences and perspectives, and elicit data librarians and information agents' new roles, this study explores the following key questions:

1. *What are scientists' experiences navigating the policy change and cultural shift in data management, sharing, and reuse?*
2. *How could data librarians and information agents help mobilize the data revolution?*

## 2. The Changing Data Landscape

In the conduct of scientific research, "the principled collection and analysis of data are vital guides for sense-making, discovery, and decision–making" (Virginia Tech Office of the Executive Vice President and Provost, 2016). The "continuity of science demands that data remain useful and meaningful over time" (American Geophysical Union, 2016b). This requires good stewardship of data and responsible management of information. When capturing data and information at work, the relationship between the data collection process and its work task context is particularly important, not only to the reliability and validity of data for original analysis, but also to the feasibility and usability of data for future reuse.

Open data access offers the opportunities to change the demographics and styles of research discovery (Leinen, 2015). Specifically, in an openly accessible and publicly available data network, the potential to link similar and re-use initially unrelated datasets would help reveal unexpected relationships and trigger new dynamics of scientific discovery (Research Data Alliance, 2014). However, the realization of these data potentials is contingent upon good organization, rich description, and reliable documentation of data. Insufficient or haphazard management practices can make study replication impossible and render data useless.

In the field of natural resources and environmental science, data, particularly forest resources data, encompass trees, other plants, animals, microbes, and environmental factors of the forest (Kang, 2001). These data provide fundamental records "about rules of dynamic changes in forest resources in terms of quantity, quality, growth and fluctuation, as well as interdependent relations between the natural environment and its operation and management" (Yan, Zhang, Jiao & Li, 2007). They provide fundamental basis for the analysis, inspection, and evaluation of the environmental ecosystem, as well as the formation of natural resource policies, with significant implications for economic development, social responsibilities, and ecological security.

In this field, scientists often deal with huge amount of data from many diverse sources that could be unstructured, highly dimensional, and containing valuable insights. To harness the implied power in data, different mining approaches and modeling systems are being developed. Often "driven and informed by taxonomies, ontologies, and other controlled vocabularies, these systems are becoming acutely able to ingest, analyze, theorize, and subsequently offer complex and dynamic solutions to many problems" (Earley, 2015). Among these applications, robust metadata models play significant roles in mining solutions by not only providing data descriptions and research information, but also building interrelationships among scientific evidences and research objects.

The scientific community increasingly recognizes the significance of storing, organizing, describing, preserving, and re-using data sets. As a result, government and private funding agencies and gradually more publishers are introducing or enforcing data management requirements and pushing for open sharing policies. The goals are to ensure long-term stewardship and accessibility of data. To address the relevant challenges in data management, previous research has studied the processes of scientific data work and promoted the engagement of adaptive curation practices and digital library systems early in the data lifecycle (e.g. Borgman et al., 2007; Palmer et al., 2012). But questions remain to be answered, for examples, how do researchers act upon funding mandates and navigate the changing practices? What are researchers' experiences, performances, struggles and concerns when working with constantly emerging and ever-growing data archives and repository platforms? Answers to these questions are critical to the strategic thinking and design process related to data policy refinement, system development, and function enhancement going forward.

In such context, this paper explores scientists' practical considerations and actual concerns when dealing with data as they strive to responsibly solve large-scale scientific challenges. It also demonstrates practical data management implementation barriers and operational deficiency in today's changing environment. Based on these findings, this study provides suggestions on repository strategies and curation pathways. It also discusses libraries' new opportunities to ensure the long-term viability of data.

## 3. Research Methods

This research has an empirical focus and was conducted through face-to-face interviews, following a snowball sampling approach as needed. The data collection includes a focus group interview with the team of scientists in the Center for Natural Resources Assessment and Decision Support (CeNRADs) and multiple individual interviews with other faculty scientists in the College of Natural Resources and Environment at Virginia Tech. A total of six natural resources and environmental scientists were interviewed at the end of 2015 and over the spring of 2016. This qualitative sample size permits the deep-dive, focused, and case-oriented analysis of a selected group of subjects. It meets the qualitative research design and sampling recommendations of Creswell (2007 & 2013) and Yin (1994).

Each interview lasted from 1-2 hours. All interviews were semi-structured by employing a newly designed protocol that emphasizes critical incidents (Flanagan, 1954), scenario building, and story telling of researchers. The use of qualitative interview method is appropriate for yielding rich descriptive accounts, interpreting and integrating multiple perspectives, and promoting holistic description that are effective for contextualized understanding of processes, practices and scenarios. The interview questions covered the breadth and depth of their scientific data work by exploring the scientists' all-around perspectives as data producers, users, managers, and providers, as well as educators. As with qualitative research methods, this project is designed to focus on practical issues, contexts, and challenges and does not purport to be numerically intended or statistically representative.

Throughout the interviews, the Principal Investigator (PI) wrote memos to record insights and reflections which further informed the following interview process and supported the development of coding framework and in-depth analysis. All interviews were fully recorded and manually transcribed, and the PI conducted open coding and axial coding for qualitative analysis and research insights. Applying grounded analysis and inductive reasoning approach, the PI analyzed the results from the coding to identify patterns, articulate emergent themes, and interpret answers to stated questions.

As results, this current study illustrates the scientists' own data adventures featuring their struggles, concerns, and difficulties, as well as their sense of value, passions, and aspirations when working with data. Using real-work examples and actual field experiences, this study offers rich detail to extract insights, uncover social and technical nuances, and elicit value judgments that are essential to the understanding of a socio-technical phenomenon such as scientific data work.

The results are meaningful because Virginia Tech's College of Natural Resources and Environment has been consistently top-ranked in the nation according to USA Today (2015; 2016; 2017). The research participants from the college have been actively involved in forest resources and environmental conservation work and conduct research in areas such as ecosystem science and management, geospatial analysis and biometrics, and natural resources planning and management. They work across a broad range of scales from individual trees to forest inventory to global earth system modeling. Uniquely positioned within the college, the Center for Natural Resources Assessment and Decision Support conducts multidisciplinary and translational work of assessing the complex dynamics of changing land use, resource conditions, ecosystem services, policy implications, and business decision-making (The Virginia Tech Center for Natural Resources Assessment and Decision Support, 2015). Situated in such context, the research participants thus provide an exemplary sample for understanding the data work critical to solving many of the world's pressing issues in resource management and environmental conservation.

## 4. Findings and Discussion

### 4.1. Provenance Tracking and Algorithm Transparency

When asked about major challenges in data search and use, the participants discussed data documentation deficiency in existing databases and archives. They particularly noted that neither data nor the contexts in which they circulate nor the mutually constitutive relationships between data and their contexts are well documented. Provenance tracking is especially troublesome in today's data preservation. As the respondents pointed out, the shortage of preprocessing information and the lack of provenance tracking are a prevalent and significant problem spanning existing databases, including public agencies' sites. As a result, research replication becomes impossible and long time-series studies are being deterred. To address this challenge, more systemic and responsible approaches, from federal policy push to community engagement, are eagerly needed. The following statements from the respondents illustrate these points.

> *"One of the big tricks that we have is, even with public agencies' [databases], the data have a provenance. So we might download and do all this preprocessing, create a long time series, and then find out that they've changed their processing to better update some new development, which means that we've done all this work on a dataset then itself has inherently changed. Then there's the question of what you then save. If you really want other researchers to work with the data at some level, the concept of 'just go download it from a public server' doesn't always work, because we may have preprocessed it significantly or [public databases] may have changed their own provenance and not have previous versions of the data available that we actually used."*

> *"What I'd like to see is broader than just Virginia Tech, really our community has to come up with a mechanism by which we can replicate the preprocessing phase, such that people will re-download publicly available data and get to the same point that we started, without necessarily having every repository storing all these data. There need to be a big federal push, because it's really quite a problem especially as [agency sites] change their own provenance and don't keep back copies. That is enormously frustrating with larger data sets, cause ideally if we've transformed [a dataset] to the point where it works well, we should be able to get any new data set and transform in the same way so that it matches the old one to extended time series, [so we can] look [at] changes across time. Sometimes it's really very difficult, so that's one of our big problems."*

In fact, poor documentation not only introduces quality assurance problems, but can also result in data being useless and discarded. To conduct research, scientists would have to produce or reproduce raw measurements, which is often quite costly. As shown below, the interviewees pointed out specific problems, including both the insufficiency of data collecting and handling information and the absence of original assumptions and analyzing details.

> *"Another specific thing that we've run into a lot, which is pretty challenging actually, is when we obtained data from legacy research, we don't always know all the specific steps that the researchers took in obtaining the numbers, [and] the measurements that they made. We have a general idea because it's written in the methods or in the study plan or in publications. But sometimes [original researchers] would make certain assumptions or calculations where they maybe calculated a mean from a number of observations, and then rather than recording all the original observations the way they were measured, they would simply record the mean for each one."*

> *"When we examine the data, sometimes very carefully, we'll see these perfect curves, just data points along a perfect curve, this tells us that the original researchers did not record the raw data, instead they made a summary model, like a regression model or something, and only recorded the predictions from the regression model. If this is the case, we'll delete those, we won't include those in our data set."*

The respondents' accounts reveal great frustration and significant cost associated with poor documentation practices at a large scale. One respondent clearly indicated the importance of discoverability and understandability of data, while also expressed concern over software preservation and made an argument for transparent data solutions.

> *"I think DataCite and the ability to actually cite data has value, but in the end the main thing is discoverability and understandability. We're not always able to do that, a lot of our processes involve some chunks of essentially non-public domain software where we have to license … So one of the questions becomes how do you allow that to be replicated, maybe you don't even know the details of the internal algorithm, all you know is that it is software version X by company Y, that's what you report in your manuscript. My big concern with that is the fact that nobody keeps old versions of the software."*

To the above points, "the vast majority of existing algorithms are opaque – that is, the internal algorithmic mechanics are not transparent in that they produce output without making it clear how they have done so" (Springer Book Overview, 2016). Such situation calls for "transparent data algorithms." Transparent data solutions are needed to produce more credible and reliable information and applications. They play a key role in proactive user engagement by making the data analytical process and its modeling details widely accessible.

Above all, when a scientific claim is published, it is essential that the evidentiary data, the related metadata that permit their access and re-analysis, the context that maximizes their semantic values, and the codes that are used in computer manipulation are made concurrently open to scrutiny (Science International, 2015). Open data stimulates data quality assurance practice and motivates scientists to maintain the vital process of self-correction.

## 4.2. Key Integrative Characteristics

Natural resources assessment often requires working with data from various agencies and diverse sources. One essential question is: what are the key integrative characteristics that allow data from multiple projects or collections to be brought together? Such examples include the presence of spatial, temporal, or taxonomic data structures.

According to the participants, a common coordinate system, units of measurement, spatial location, and uncertainty estimation are key ingredients for integration. In addition, compatible classification systems, sufficient metadata documentation, and consistent categorical structures are also key integrative characteristics for data instrumentation. But often such information is missing, inconsistent, or inaccurate, which causes major obstacles for research performance. In particular, the lack of uncertainty estimation and units of measurement frequently impede the collection, use, and reuse of existing data sources. In many cases, "even when the data are found, information on units, protocols, and column names are garbled, missing, or inaccurate" (Desai, 2016). Multiple respondents articulated several examples below.

> *"It can be something as simple as a common coordinate system. That's actually sometimes an issue for us to make sure everybody is using exact same datum, exact same coordinate system, and these things can actually be related one to the another. Units, really important, can't have somebody doing one metric and another doing something else. I think the lack of units, even [by] an established scientist, is a huge ridiculous problem."*

> *"Spatial location is the key thing that integrates data ... Beyond that, I'd say classification systems have to be compatible ... there has to be categorical consistency."*

> *"Obviously good metadata ... Increasingly the work I'm doing actually benefits from uncertainty estimation that travels alongside the observations ... I want to know the standard deviation of the distribution around a particular measurement, or the 95% confidence intervals, variance, or the resolution ... There's a scale from raw data to data products, which mean there's been some sort of model applied to it ... But one of the issues is actually these data products not having uncertainty estimation, which makes it hard."*

These comments resonate with the points enumerated in the Federal Big Data Research and Development Strategic Plan (Networking and Information Technology Research and Development Big Data Senior Steering Group, 2016). In particular, to ensure the trustworthiness of information and knowledge derived from Big Data, appropriate methods and approaches are needed to capture uncertainty in data and to ensure reproducibility and replicability of results. This is vital especially when data are repurposed for a different use and when data are integrated from multiple, heterogeneous sources of various qualities.

Another important condition for conducting data integration is the actual "physical [storage and computing] space" for Big Data computing. As one respondent indicated "working with terabyte of data, but having them in multiple spots, for a lot of our workflows, can be problematic." This revealed the important data and workflow management challenges in distributed, parallel, or cloud computing.

### 4.3. Data Management Concerns and Struggles

The participants also expressed their own data documentation, archiving, and preservation problems, concerns, and struggles. They not only admitted their own documentation crises, but also echoed to the fact that "well-managed data foster good scientific practice, and sharing well-formed, well-described data increases research impact" (OpenAIRE & EUDAT, 2016).

> *"Mostly we just make our own readme files ... they aren't really true metadata standards. The problem is keeping the darn things together. Sometimes we have readme files but have no idea what data sets they actually go with ... We need much better naming, folder creation, and archiving, the whole thing just needs to be better."*

> *"We don't do enough of that. We've been talking about the need to document our methodology and approach, and we really need to document our output data. We value metadata and source data. I really appreciate when we look for data and it has rich metadata. But we have not provided a lot of that on our end, for our outputs."*

From the scientists' points of view, documentation is not just for other people's understanding but, most importantly, for their own understanding and research effectiveness.

> *"All these scenarios that we are looking at and running, for every set of output data, we need to make sure we know what all the input assumptions are ... right now our documentation schema is to read my notebooks. Each [team member] just takes [his/her own] notes about what we do, we don't have a [standardized practice]. Like the scenarios, I got confused about what the assumptions of one scenario versus the other were. So it's a challenge. It will be easier for us to develop a template to do some standardized documentation."*

A participant also voiced concern on maintaining data in an archival format that will be readable in the future. In his opinion, simple, open-sourced data formats are preferable.

> *"Whereas remotely sensed data, we got immense huge files that have to be stored, and then there's always a question of: are you maintaining in an archival format that will be readable in the future? I am very worried about that because honestly if space was not a concern, I think we should keep it all in ASCII, because it might take 6 to 10 times more space [but] there would never be an issue of being able to access ... you can read it, you wouldn't have to worry about some third-party software."*

Another participant cautioned against problems such as storage media degrading, files getting corrupted, and data formats becoming obsolete (Monastersky, 2013).

> *"So long-term sustainability, there is no guarantee. I am kind of cautious about it because I have seen a lot of data on magnetic tape, IBM tape, cassettes, or on mainframe punch cards never were saved. Hopefully people will be pretty considerate about preserving these things."*

Long tail data, including field-based data and smaller spatial acquisitions, are where scientists have the most trouble with file naming and metadata documentation. For example, when asked if metadata standards and documentation schema are used, a researcher commented as below.

> *"To be honest, with our field-based data, I would say most of the time we do a horrible job with that ... like some little CSV files where you have 1000 of them and are trying to figure out which one is the right one. You didn't maintain a common naming system through time. It is actually the field data that we have the worst trouble with. Once in a while, some of these smaller spatial acquisitions also cause trouble. I got a LIDAR data set from 1999, at the time I couldn't figure out where it was, who's done what processing, I was just lost ... Usually the older it is, the worse it is."*

To sum up, scientists need to learn how to develop data and metadata management workflows that support the preservation and discovery of these resources.

## 4.4. Data Policy Compliance and Operational Deficiency

When asked what can help with better data documentation, the participants indicated that data management and sharing should go far beyond complying with funding commitments. They believed that researchers should take up a more proactive role by recognizing the values of maintaining well-managed data and the benefits of sharing well-described data. They also pointed out that the existing National Science Foundation or other federal agencies' data sharing mandates are not effectively producing the anticipated outcomes of supporting actual reuse or future discovery.  Data policies thus need to be revised to better reflect and align with the value propositions and intended objectives.

> *"What I'd like to see is more examples of [good data management practice] where it's not funded and required by the sponsor ...  where it doesn't necessarily happen that way."*

> *"One thing I think is pretty interesting is the federal requirement to preserve data ... they have requirements for data management, reporting, and sharing. [But] they are lacking [in*

> *performance], primarily because researchers see it as a requirement, so they make the efforts to meet the requirement, but they don't actually in many cases preserve data in the most … raw form … When I go to [a federal funding agency] websites, download the data, what we find is that they've been processed and filtered through some statistical analysis and models. Oftentimes they've taken out some of the actual normal variation that was inherent to the original measurements, and standardized the measurements, so there's no way to actually reconstruct the original measurements based on what they actually share on the public repository … So [researchers] met the requirements, but it is not supporting future discovery. That's little complicated, but it is something that hopefully people will be thinking about as they develop these policies. You know any policy has unintended consequences. That's what we see in something like NEON [National Ecosystem Observation Network] or LTER [Long Term Ecological Research], they meant very well but unfortunately the results haven't lived up to the best possible outcomes."*

In addition to policy adjustments, researchers' own sense of value in data sharing to support greater opportunities for discovery plays a key role.

> *"I honestly think it's still going to require some degree of recognition by the individual contributors that there's some value in it for them, and I think most of them do have a sense of some value to them that other people will use their data, and it can be part of something bigger, maybe some greater discovery, or greater opportunities for discovery, and also some recognition."*

> *" … I think this is ethical value that most researchers carry through their careers …  [What is really helpful is] researchers have instilled into their graduate students, into the next generation, inherent value of data."*

The participants' comments reveal the importance of catalyzing a systemic cultural change, from governmental push to individual commitment, with institutions standing in between to grease the wheel and strike a balance.  In particular, with research funders' much-discussed data expectations taking effect, institutional data management services, many of which are library-hosted, are "often charged with aiding and monitoring policy compliance" while also trying to keep everyone happy. Compliance is important, but obviously not self-sufficient. Institutions need to "look not only at the delicate balance between compliance and cooperation, but also at the wider benefits to universities and researchers of a pro-active approach to data stewardship" (Research Data Management Forum, 2016).

### 4.5. Data Loss and Missed Opportunities

According to the participants, some research opportunities may have been shelved due to poor data stewardship or preservation practices. One researcher's own encounter demonstrates that data loss means lost opportunities in discovery.

> *"We know there were some data sets, very rich data sets, that were lost. We could have better models, better estimates of different attributes of forests, biomass, carbon, nitrogen,* *and whatever elements are for forest trees. But because those data were lost we don't have that opportunity to discover."*

The same participant also enlisted specific incidents where data were accidentally lost, willfully destroyed, recklessly tossed, or carelessly forgotten, and when data got missing in physical transition or human transfer. These various cases of data loss are quite alarming and offer all the more reason for responsible and rigorous management and preservation of data.

*" In the 1970s or 1980s, a lot of this kind of work was done in that era. In the US, we had the energy crisis in the 1970s, many different universities and federal agencies were pursuing the same kind of work with the felled tree measurements. This one researcher collected a lot of data, made many publications, he was probably among the most active researchers in this area, but when he left the position … he destroyed all of the data, it's never been found. That's one case where we found the data were either accidentally lost or willfully destroyed. The other is sometimes institutions or individuals possess the data, they're not willing to share it for various reasons … I guess another one is really just time, too much time has gone by and the data disappeared, no record of where they were … maybe the paper files were lost or thrown out. Today the problem still exists, sometimes it's on a person's computer and then they get a new one but forget to put the old files on the new computer, because now they have moved on to their next project. We called this professor at the University of … and said, 'in 1993 you made a publication and you had 70 trees measured, do you still have the data?' And they said, 'oh, it got lost.' Sometimes we find, sometimes not."*

According to another participant, contracted, licensed, and restricted data sets are considered impeding research workflow because of the difficulty to share and reproduce.

*"Sometimes we contract with a third party to acquire data using a sensor that a company has, in which case there may be licensing restrictions … you cannot then re-share them with anyone else, and that's true for some other satellite-based products as well … We increasingly try not to do that because it impedes our workflow, to be honest, and it makes the ability to truly reproduce results by another research group more difficult."*

In sum, scholars' agendas are often dictated by the data and information that, by happenstance or design, are readily available. This is exemplified in a participant's comment below.

*"How much of forest is owned in small parcels [is] one of the key things in forestry. If I'm a landowner, I have 5 acres of forest, it's unlikely it will ever be harvested for wood products, [because] it's just too small. If I have hundreds acres of forest, it's highly likely. So this parcel size is important to know … While counties have been slow in sharing that data, it is valuable for research purposes. But it is still incomplete, even now in Virginia about 130 counties, there are 15 or 20 counties that have yet to share their data. So we cannot claim that we've done this work. We could easily do the statewide analysis if the data were available. But it's not. So we can't claim to have a statewide answer."*

## 4.6. Data Models and Standards

It is widely recognized that data models are often developed to help researchers conceptualize, act on, and reason about data in a more atomic and structured fashion. To this end, existing taxonomies, ontologies, and other controlled vocabularies are used to help inform this process and deepen the analysis of unstructured data. But on the other hand, these existing schemes and standards may have also prohibited the processes of data discovery and analysis, especially those of novel and innovative nature. One participant demonstrated this conflict.

> *"For instance, I have a colleague at … University, his measurement standards and protocols are much different than almost anyone else. When he measures felled trees, he records a lot more information … not only the standard variables but also many additional variables that are not part of the standards or not part of the textbooks from the past. He hopes to discover something else, maybe set new standard going forward, that would be exciting. But on the other hand, it really does help when people have followed the same standard, because without that we wouldn't be able to tie our data together … These [new, additional measures] would be [useful for the future], absolutely, he has convinced me that they would be. But there's nothing we can do about what has been done in the past. They [the existing standards and models] can be in some sense a barrier, pros and cons, no doubt about it."*

It has been well recognized that "knowledge organization and representation activities contribute greatly to the sustainability and long-term success of a research data curation system." These activities co-evolve with the discipline or domain that they serve. We need to understand how they evolve and how data models and metadata schemas should be edited and revised to "accommodate changes in scale, complexity, or heterogeneity of research data" (Data Science Journal, 2015).

## 4.7. Data Curation Values

Next, the participants were asked to describe what data are particularly valuable to curate for long-term reuse and repurpose. They emphasized the importance of preserving longitudinal data collections, not only to ensure the rigor of research and support future projections, but also to ensure data vitality over time through cross-examination or inter-calibration across different time series.

> *"The most important data sets are the ones that are uniform across space and time and repeated. It's really the longitudinal studies across time using the same measurement protocols with large number of participants that have really high rigor. So for example, we're about to do a fourth or fifth re-measurement of the set of plots in the state forests, and every time we pay to re-measure that, [the data set] becomes more and more valuable because it is starting to give you a feel for conditions through time, [researchers] start to ask a variety of questions [and] pay attention to metric inter-calibration and quality across time."*

> *"We are doing future projections that really can't be substantiated. But we could and*

> *probably will, do a retrospective analysis where we go back to the year of 2000 and project forward, and hopefully use that to ensure the quality of our models and assumptions as much as the data … If we have this long enough when going back using older data and running these projections to current day, we should be able to see if these projections are correct."*

According to the participants, there is no definitive answer as to whether raw data or processed data are particularly valuable for long-term use. This is because the same data set could have unpredicted values for new purposes, or could work under various scenarios to answer different types of questions within and across disciplines. A few examples from multiple respondents are shown below.

> *" To the point where we were using remote sensing-based forest patches, we already knew there's a lot of value that we will glean from the simulation runs. So a quick example, once we have a map of all these forests in Virginia in age-class distribution, and start simulating these models … then we basically have forest cover maps for the state for 30 years into the future under a variety of scenarios and multiple iterations. Then there are other researchers [who find these useful]. I was talking to a wildlife professor who said if I had a map like that, it could predict quail habitat 30 years into the future using land cover maps with just what we're producing. So then we would need to preserve not only raw outputs but basically a map at each iteration of the model and outputs of plot data in inventory every year, because those data could be fed into other models, [such as] wildlife habitat or water quality models, etc."*

> *"Does it make more sense to store the raw or the summarized data? When we summarize it, we're implying that we know what should be summarized on. Maybe somebody else would have a different [goal] … I don't know. So for example, as to the model outputs, we could curate the raw for the first level summarization, then simulated landscapes, and plot inventory data, things like that, [it] would be a big quantity. We have not given that a lot of thought yet."*

> *"Our philosophy is to preserve raw data … A regression model is a summary of data, but the raw data themselves convey the full information of what was actually observed when the data were collected. So we generally try not to reproduce model predictions, but produce or reproduce only actual raw measurements. If people want to make adjustments or average them later, they can do that on their own."*

Above all, how to approach the complexities of raw data versus aggregated data remains a difficult question. Generally, scientists benefit from access to raw data, which can always be used to feed into research records or computational tools to generate aggregation and analysis. Aggregated data can be appropriate for reuse only if proper provenance tracking, sufficient research documentation, and uncertainty estimation information are well maintained.

## 4.8. Libraries' Roles in Data

Speaking of libraries' roles in data stewardship, one participant highly valued the dedicated efforts and professional support of libraries to curate, archive, and preserve their data in a sustainable and responsible fashion.

> *"I am encouraged and excited to hear about the opportunities for curating, archiving, and storing the data that we produce … I think it's really important, because philosophically I am old enough now to start seeing the problems we have because we haven't done that for the last 30 years."*

When asked how libraries could make an impact, another respondent used a familiar analogy. In his view, checking out data sets should be just like checking out textbooks or any other library resources.

> *" If I am an active computer, I don't mean that as a physical device but as a person, let's say I am teaching a class on X or Y, I should be able to essentially check out [data sets] just like I would over textbooks on reserves. I should [be able to] check out data sets and utilize as a student, and work within that context in the same way as I would with any library resources."*

This analogy touches on data access issues.  As more scientific and scholarly interactions with libraries happen digitally, interfaces and tools for access have become increasingly important. Thus, more work on user participatory design and research at the human-data interface are necessary to optimize the usability and accessibility of data libraries.

Interestingly, previous research surveyed academic research libraries (Tenopir et al., 2014) and science librarians (Antell et al., 2014) regarding their awareness and perceptions, involvement and readiness, roles and responsibilities in research data management. Majorly focused on data management planning and consultations, or further, data archiving and preservation, libraries' roles in data are limited in scope. In contrast, the identification of scientists' and scholars' perspectives and needs has shown a much broader horizon in data support and service requirements, with a different emphasis on data mobilization and workflow efficiency in aspects such as processing, manipulation, and utilization (Shen, 2015; 2016; 2017a; 2017b). Adding another dimension, the scientists in this study expect the integration of data libraries into the fabric of researchers' workflows with seamless data interface and anywhere accessibility. Such insights are particularly valuable for digital libraries to envision their future direction and strategize their investment in efforts to generate broader impact.

## 4.9. Data Skills for Future Professions

Finally, one participant envisioned that the fundamental skills of managing, interpreting, and analyzing data would be required in almost every future profession. The workforces of present and future generations will need to demonstrate good problem solving skills with the ability to both interpret problems and apply appropriate data solutions in any field they serve.

> *"In almost any profession, starting now in their careers, people will have opportunities and needs for working with data and delivering solutions related to data. Data really just permeates almost every major profession, including forestry. So I think it is pretty important to have young people who are dedicated to their professions to have a solid foundation for data science ... so that they can put it in the context of how they can use it. Just like being able to write and communicate with people, these are fundamental skills: being able to manage data, interpret it, and understand what it can do for you and how."*

To sum up, a fundamental change of the dominant culture starts with appropriate training in comprehensive data skills that should be embedded in educational experiences and workforce development. The development of these basic skills and competencies should become required components and integrated norms in higher education system and shouldn't be dominated or steered by research funding or policy incentives.

# 5. Conclusions

## 5.1. Data Quality Assurance

To conclude from the current research findings, several major themes emerge. Firs of all, data quality assurance is fundamental to any scientific endeavor. As such, the American Geophysical Union is calling for better management of earth science data quality information for the benefits of operational use and community re-use. Here, scientific data quality is defined in terms of "accuracy, precision, uncertainty, validity and suitability for use" (American Geophysical Union, 2016a).

In scientific fieldwork, many data samples "have been collected at great cost and with substantial difficulty." They can be "rare or unique, and irreplaceable" (The First International Physical Samples and Digital Libraries workshop, 2016). In such cases, scientists cannot afford data loss or poor quality. We recognize that there are costs in preserving data for reuse, but there are often even greater costs in not doing so, particularly when research is irreproducible. There is no amount of funding that can reconstruct ephemeral or time-dependent phenomena for which the data were not well preserved and properly curated (McNutt, 2016). Thus, scientific communities must exercise robust stewardship activities.

According to the current research participants, sharing data can actually act as an effective quality checking mechanism by allowing re-users to detect unintentional errors. "The expectation that others may be verifying data could also encourage greater care in original analyses" (McNutt, 2016) and trigger vigilant self-correction data practices.

## 5.2. Data Preservation Values

To ensure that observations, experiments, and models are reproducible often requires access to original data (Leinen, 2015), which challenges all data repositories to accommodate and curate many more data sets. However, aggregated, derived, or processed data products can also prove useful for replication and repurpose when effective provenance tracking, sufficient documentation, and algorithmic transparency are in place.

According to the participants, historical, longitudinal data collections that have been consistently documented are particularly valuable for preservation and curation. This is particularly true because these types of data contribute to environmental monitoring and climate surveillance over time. Increasingly, researchers are going beyond the analysis of extant data and starting to work with data archivists and information scientists to perform historical or longitudinal time-series investigations. Data librarians have the opportunities to support data instrumentation and revolution in this process.

To do so, data librarians are making efforts to assure the availability and accessibility of a more systematic collection of data for analysis, synthesis, and critique. They are also committed to assuring the accuracy and consistency of information representation and knowledge documentation. These efforts are all made in hope to unlock the semantic black box of data for accountable repurpose. As scientists' research agendas are often dictated by the data and information that, by chance or design, are readily available, data librarians need to find solutions to ensure data discoverability, algorithm transparency, and holistic curation. Since "open data access has the ability to change the demographics of research and the styles of discovery" (Leinen, 2015), data libraries have the potential to increase the efficiency and capability of scientific data work to address grand challenges and create societal, environmental and economical benefits.

### 5.3. Reuse Struggles and Metadata Framework

Today, scientific communities increasingly recognize the immense significance of storing, discovering, processing, preserving, and re-using data sets, workflows, and software (Metadata and Semantics Research Conference, 2016). However, extracting knowledge from data is still a daunting task. For one, data sources are not integrated. Some contain private information and are not structured. Many suffer inconsistent management, incomplete storage, or insufficient documentation. For another, many data sources lack context-sensitive and privacy-aware algorithms to support robust inquiry and effective reuse. Data made publicly available in compliance with governmental policies or funding mandates are presumably valuable for future discovery, although not always in forms that enable reuse or support analysis.

Metadata is a critical mediator in this process, providing the means to render black-box digital files discoverable and reusable. To such effects, rich metadata about research outputs needs to be recorded and disseminated, including contextual, semantic, and provenance information. Yet, the recording and utilization of domain-specific information is always evolving. Metadata framework not only needs to represent the ever-developing domain knowledge. For certain use cases, metadata also needs to be uniformly accessed across research domains, to foster collaboration and exchange among different disciplines and vertical communities (Metadata and Semantics Research Conference, 2016).

To capture the development of a field, we need to provide a forum for domain scientists to reflect on state-of-the-art metadata and semantic evolutions. This would help inform open repositories, research information systems, and data infrastructures development (Metadata and Semantics Research Conference, 2016). More research is also needed on improving the trustworthiness and usability of data (Networking and Information Technology Research and Development Big Data Senior Steering Group, 2016).

## 5.4. Data Modeling and Knowledge Representation

In the context of scientific research, "as large datasets are mined, analyzed, and clustered, the key properties, relationships, categories, and structure in the data are transformed into knowledge" (Research Data Alliance, 2015). Knowledge organization and representation activities co-evolve with the discipline or domain they serve, and play significantly in the success of a research data curation system. Today's research landscape features great acceleration of scientific discovery and constant evolution of knowledge structure. In this landscape, we need to keep defining, evaluating, curating, refining, and maintaining taxonomies and ontologies, while "navigating structured information and search strategies for complex data" (Research Data Alliance, 2015).

Future research could also pursue data-driven ontology and schema development, which could then be compared, contrasted, and integrated with expert domain knowledge to vet, refine, and evolve data standards and drive the knowledge discovery forward.

## 5.5. Data Curation as Creative Scholarship

Above all, curation activities are crucial for data-intensive research in natural resources and environmental science, and equally at the heart of understanding social and economic sustainability. In particular, the curation of natural resource data into electronic databases offers opportunities to better understand and predict environmental ecosystems and human impacts. The capacity of digital resources to endure is also a key factor in such undertakings. To preserve the long-term vitality of research assets in the field, we need to leverage existing archival data, optimize information processing, and strategize vocabulary reconciliation and metadata remediation. In addition, data services with aggregating capability can become a new territory for data librarians to unleash creativity and build scholarship. Such task requires an expert who is fluid and fluent with data to merge research assets for scientists.

After all, data curation can be a space for deep learning and knowledge discovery. It can be an intellectual driver to connect disciplines and inspire creativity. Data curation can become scholarship, a new form of creative scholarship, designed to serve a diversity of research experiences and expectations. In such endeavors, data curators will be challenged to become problem solvers, change agents, and creative scholars to address the increasing complexity of the world.

## 5.6 Future Direction

This empirical socio-technical study of natural resources and environmental scientists' data work reports the practices, pitfalls, requirements, and desiderata of data management systems in this field. The results provide implications and recommendations for scientists, librarians, developers of data management systems, and higher education administrators on ways to improve data collection, use, and reuse in the sciences. This study resurfaces some long-standing issues such as the needs to adopt a common vocabulary for data-related practices; the need to record algorithms used to process, analyze, and summarize the data; and the need to track provenance and document "normal" irregularities in data. The results also reveal new issues such as data policy compliance problems and operational deficiency, storage and computing space issues for big data processing, as well as data and workflow management challenges in distributed, parallel, and cloud computing.  All these issues need to be taken into consideration in the design and construction of data management platforms.

Further work should delve into scientists' actual data sets and their specific data gathering and curation actions. More research could be done to contextualize or contrast the expectations, actions, and proficiencies of scientists' data work from different fields, or from the same field at a different time.

Among researchers or practitioners who have been focusing on data-related practices in the sciences, there has been a constant voice calling to change the dominant culture and set up appropriate incentives to really create needed actions. But maybe it is time to jump out of this echo chamber and entrenched vision to join the real action of culture change. As higher education landscape is transforming so rapidly to embrace data sciences in almost every domain or cross-domain strategic areas, information science scholars and data practices researchers need to bring data revolution into movement and set agenda for integrating the development of comprehensive data skills into the academic norms of higher education.

## Ethics and Consent

All appropriate human subjects procedures were approved by the Virginia Tech Institution Review Board (IRB) and followed under VT IRB-15-968.

## References

American Geophysical Union (2016a). *The 2016 AGU Fall Meeting session description: Managing earth science data quality information for the benefit of users*. Available at https://agu.confex.com/agu/fm16/preliminaryview.cgi/Session12500 [Last accessed 15 September 2016].

American Geophysical Union (2016b). *The 2016 AGU Fall Meeting session description: Publishing and managing data: The case for trustworthy digital repositories*. Available at https://agu.confex.com/agu/fm16/preliminaryview.cgi/Session13445  [Last accessed 1 September 2016].

Antell, K., Foote, J. B., Turner, J. and Shults, B. (2014). Dealing with data: Science librarians' participation in data management at Association of Research Libraries institutions. *College & Research Libraries* 75(4): 557-574, DOI: https://doi.org/10.5860/crl.75.4.557

Borgman, C. L., Wallis, J. C. and Enyedy, N. (2007). Little science confronts the data deluge: Habitat ecology, embedded sensor networks, and digital libraries. *International Journal on Digital Libraries* 7(1-2): 17-30, DOI: https://doi.org/10.1007/s00799-007-0022-9

Creswell, J. W. (2007). *Qualitative inquiry and research design: Choosing among five approaches*. Thousand Oaks: Sage Publications.

Creswell, J. W. (2013). *Qualitative inquiry and research design: Choosing among five approaches.* London: Sage Publications.

Data Science Journal (2015). *Special issue introduction: Advances in data modeling and knowledge representation for research data*. Available at http://www.codata.org/news/94/62/Data-Science-Journal-Call-for-Papers-Advances-in-Data-Modeling-and-Knowledge-Representation-for-Research-Data [Last accessed 1 September 2016].

Desai, A. R. (2016). Your science is your (openly shared) data. *Eos Earth & Space Science News*, 26 May. Available at https://eos.org/editors-vox/your-science-is-your-openly-shared-data [Last accessed 1 September 2016].

Earley, S. (2015). Cluster analysis in big data mining explained - Without the math. In: *Earley Information Science Blog*, September. Available at http://www.earley.com/blog/cluster-analysis-big-data-mining-explained-without-math [Last accessed 1 September 2016].

Flanagan, J. C. (1954). The critical incident technique. *Psychological Bulletin* 51(4): 327–359.

International Data Week (2016). *SciDataCon 2016 session: Defining data professionals.* Available at http://www.scidatacon.org/2016/sessions/98/ [Last accessed 15 September 2016].

Kang, X. (2001). *Forest management*. Beijing: Chinese Forestry Publishing House.

Leinen, M. (2015). The data flood: Implications for data stewardship and the culture of discovery. In: *DataOne Webinars*, 8 September. Available at https://www.dataone.org/webinars/data-flood-implications-data-stewardship-and-culture-discovery [Last accessed 1 September 2016].

McNutt, M. (2016). #IAmAResearchParasite. *Science,* 351(6277): 1005. DOI: http://doi.org/10.1126/science.aaf4701

Metadata and Semantics Research Conference (2016). *Special Track: Metadata & semantics for open repositories, research information systems and data infrastructures.* Available at http://www.mtsr-conf.org/index.php/tracks/open-repositories [Last accessed 1 September 2016].

Monastersky, R. (2013). Publishing frontiers: the library reboot. *Nature,* 495 (7442): 430–432. DOI: http://doi.org/ 10.1038/495430a

Networking and Information Technology Research and Development (NITRD) Big Data Senior Steering Group (2016) *The federal big data research and development strategic plan.* Available at https://www.nitrd.gov/PUBS/bigdatardstrategicplan.pdf [Last accessed 15 September 2016].

OpenAIRE and EUDAT (2016) *Research data management: An introductory webinar.* Available at https://eudat.eu/events/webinar/research-data-management-an-introductory-webinar-from-openaire-and-eudat [Last accessed 1 September 2016].

Palmer, C. L. et al. (2012). *Site-Based Data Curation at Yellowstone National Park.* Available at http://cirss.ischool.illinois.edu/SBDC/index.php [Last accessed 22 September 2017].

Research Data Alliance (2014). *Rome recommendations on the future of research data and computing infrastructures.* Available at https://rd-alliance.org/sixth-plenary/e-infrastructures-rda-data-intensive-science.html [Last accessed 1 September 2016].

Research Data Alliance (2015). *Pre-RDA plenary workshop: Infrastructure for understanding the human rain, Session 3: Knowledge management and search*. Available at https://rd-alliance.org/plenary-meetings/sixth-plenary/programme/e-infrastructures-rda-data-intensive-science/infrastructure [Last accessed 1 September 2016].

Research Data Management Forum (RDMF) (2016). *RDMF15: The compliance of science? Data policies, expectations and concordat.* Available at http://www.dcc.ac.uk/events/research-data-management-forum-rdmf/rdmf15-the-compliance-of-science [Last accessed 1 September 2016].

Science International (2015). *Open Data in a Big Data World.* Paris: International Council for Science (ICSU), International Social Science Council (ISSC), The World Academy of Sciences (TWAS), InterAcademy Partnership (IAP). Available at http://www.icsu.org/science-international/accord/open-data-in-a-big-data-world-long [Last accessed 1 September 2016].

Shen, Y. (2015). Research data sharing and reuse practices of academic faculty researchers: A study of Virginia Tech data landscape. *International Journal of Digital Curation* 10(2): 157-175, DOI: http://dx.doi.org/10.2218/ijdc.v10i2.359

Shen, Y. (2016). Strategic planning for a data-driven, shared-access research enterprise: Virginia Tech research data assessment and landscape study. *College & Research Libraries* 77(4): 500-519, DOI: http://dx.doi.org/10.5860/crl.77.4.500

Shen, Y. (2017a). Data sharing practices, information exchange behaviors, and knowledge discovery dynamics: A study of natural resources and environmental scientists. *Environmental Systems Research* 6(1): 1-14, DOI: http://dx.doi.org/10.1186/s40068-017-0086-5

Shen, Y. (2017b). Burgeoning data repository systems, characteristics, and development strategies: Insights of natural resources and environmental scientists. *Data and Information Management* 1(2): 1-9, DOI: https://doi.org/10.1515/dim-2017-0009

Springer Book Overview (2016) *Towards transparent data mining for Big and Small Data*. Available at http://dbdmg.polito.it/glass-boxDM/ [Last accessed 1 September 2016].

Tenopir, C., Sandusky, R. J., Allard, S. and Birch, B. (2014). Research data management services in academic research libraries and perceptions of librarians. *Library & Information Science Research* 36(2): 84-90, DOI: https://doi.org/10.1016/j.lisr.2013.11.003

The First International Physical Samples and Digital Libraries Workshop (2016). Physical samples and digital libraries, Newark, NJ, USA, 22-23 June 2016. Available at http://www.earthcube.org/announcements/physical-samples-digital-libraries-workshop-jcdl-june-22-23-2016-newark-nj [Last accessed 15 September 2016].

The Virginia Tech Center for Natural Resources Assessment and Decision Support (2015). *Research projects*. Available at http://cenrads.cnre.vt.edu/research.html [Last accessed 1 December 2015].

USA Today (2015). *The 10 best colleges for studying natural resources and conservation.* Available at http://college.usatoday.com/2015/11/20/colleges-conservation-natural-resources/ [Last accessed 4 September 2017].

USA Today (2016). *The 10 best colleges for studying natural resources and conservation.* Available at http://college.usatoday.com/2015/11/20/colleges-conservation-natural-resources/ [Last accessed 14 September 2016].

USA Today (2017). *The 10 best U.S. colleges for studying natural resources and conservation.* Available at  http://college.usatoday.com/2017/05/19/studying-natural-resources-conservation/ [Last accessed 4 September 2017].

Virginia Tech Office of the Executive Vice President and Provost (2016). *Destination area: Data and decision sciences*. Available at http://www.vt.edu/content/dam/vt_edu/academics/destination-areas/data-decision-sciences.pdf [Last accessed 1 September 2016].

Yan, P., Zhang, X., Jiao, F. and Li, S. (2007). Distributed heterogeneous forest resource data management and service architecture research. *New Zealand Journal of Agricultural Research*, 50(5): 965-973.

Yin, R. K. (1994). *Case study research: Design and methods.* 2nd ed. Thousand Oaks, CA: Sage Publications.